\documentclass[aps,pra,showpacs,preprint,superscriptaddress]{revtex4-1}
\usepackage{graphicx}
\usepackage{bm}
\usepackage{amsmath}

\begin{document}

\title{A theory addressing the quantum measurement problem: collapse occurs when the entangling speed reaches a threshold}

\author{Sang Jae Yun}
\email[]{sangjae@kias.re.kr}
\affiliation{School of Computational Sciences, Korea Institute for Advanced Study, Seoul 130-722, Korea}

\date{\today}

\begin{abstract}
To resolve the quantum measurement problem, we propose an objective collapse theory in which both the wavefunction and the process of collapse are regarded as ontologically objective. The theory, which we call the entangling-speed-threshold theory, postulates that collapse occurs when the entangling speed of a system reaches a threshold, and the collapse basis is determined so as to eliminate the entangling speed and to minimize its increasing rate. Using this theory, we provide answers to the questions of where and when collapse occurs, how the collapse basis is determined, what systems are (in other words, what the actual tensor product structure is), and what determines the observables. We also explain how deterministic classical dynamics emerges from indeterministic quantum collapse, explaining the quantum-to-classical transition. In addition, we show that the theory guarantees energy conservation to a high accuracy. We apply the theory to a macroscopic flying body such as a bullet in the air, and derive a satisfactory collapse basis that is highly localized in both position and momentum, consistent with our everyday observation. Finally, we suggest an experiment that can verify the theory.  
\end{abstract}


\maketitle

\section{Introduction}

According to the orthodox interpretation of quantum mechanics, there are two different types of processes in the universe: deterministic unitary processes and indeterministic collapse processes that occur at measurement. This dualism has made many physicists uncomfortable. Because, even at measurement, one can define a closed system that contains both the measured system and the measuring apparatus, whether the collapse actually occurs has remained controversial. Moreover, even if one accepts the collapse postulate, the question of what conditions are required to achieve measurement has remained problematic. In addition, the question of why classical objects are observed in a certain preferred basis among infinitely many legitimate bases has also been intensively discussed. These problems are collectively known as the quantum measurement problem and have been a subject of debate since the birth of quantum mechanics.

Many theories and interpretations have been proposed to address the quantum measurement problem. Some of them explicitly or at least tacitly accept the collapse postulate, whereas others reject the notion of collapse. On the collapse side, examples include the Copenhagen interpretation, the von Neumann-Wigner interpretation \cite{Wigner1967Collection}, and objective collapse theories such as the Ghirardi-Rimini-Weber theory \cite{Ghirardi1986} and the Penrose theory \cite{Penrose1996}. On the no-collapse side, examples include the de Broglie-Bohm theory \cite{Bohm1952}, the many-worlds interpretation \cite{Dewitt1970}, the many-minds interpretation \cite{Albert1988}, the consistent histories interpretation \cite{Griffiths1984}, and many others. Despite the variety of these endeavors, there has been no broad consensus that the problem was clearly solved.

The theory proposed in this paper is an objective collapse theory in which both the wavefunction and the process of collapse are regarded as ontologically objective. We accept the dualism that states that there are fundamentally two different types of processes in the universe: unitary processes and collapse processes. Because current quantum theory is satisfactory for unitary processes, we intend to establish a theory about collapse processes. The key postulate of the theory is that the state of a system collapses when the entangling speed of that system reaches a threshold. We call this theory the {\it{entangling-speed-threshold theory}}. Using this theory, we provide plausible answers to the questions of where and when collapse occurs, what determines the collapse basis, how subsystems should be defined given a large system, and what determines the observables (or, more generally, the measurement operators). We also explain how deterministic classical dynamics emerges from indeterministic quantum collapse, where nearly continuous collapse plays a crucial role in explaining the quantum-to-classical transition. In addition, we show that before and after collapse, energy is accurately conserved when the environment consists of many degrees of freedom. To convince ourselves that the theory is consistent with everyday observations of classical phenomena, we apply the theory to a macroscopic flying body such as a bullet in the air, and show that the collapse basis of the bullet derived by the theory has both a highly localized position and a well-defined momentum. The success achieved in deriving the classical states of a macroscopic body can be considered as evidence that the theory is well suited for explaining the quantum-to-classical transition. Finally, we suggest an experiment that can verify the theory.

\section{Entangling-speed-threshold theory}

Let us first sketch the main idea that leads us to the proposed theory. It seems that entanglement should be the key to solving the measurement problem because one of the essential differences between the quantum world and the classical world is the presence or absence of entanglement. If we wish to establish a theory such that collapse occurs when some physical threshold is reached, it is reasonable to expect that such a threshold should be an entanglement-related quantity. Entanglement itself is not an appropriate choice for the threshold for the following reason. If collapse occurs when entanglement reaches a high threshold, even the classical world should exhibit entanglement-related phenomena because highly entangled systems that have not yet reached the threshold would exist in the classical world. This supposition is contradictory to everyday observations. As a suitable threshold that overcomes this difficulty, we choose the entangling speed, i.e., the time derivative of the von Neumann entropy of a system. The entangling speed can be large even when entanglement itself is small, and, in particular, it can be enormous when a system is simultaneously interacting with a large number of environmental particles. Simultaneous interaction with a myriad of particles is a common feature of macroscopic classical objects. Hence, if we postulate that collapse occurs when the entangling speed reaches a certain threshold, then macroscopic objects should be able to reach that threshold easily. After collapse, the entangling speed of the object can increase again very rapidly, resulting in multiple consecutive collapses within a short period of time. We will see that this nearly continuous collapse causes macroscopic object to behave classically. In this respect, we expect the entangling-speed-threshold theory to suitably explain the quantum-to-classical transition.

\subsection{Where and when collapse occurs}

The following question is important in the foundation of quantum theory: given a large system, how is the {\it{actual}} tensor product structure determined? Or, put another way, what are the systems? This question plays a crucial role in all discussions of the measurement problem and is a particular focus of the decoherence program \cite{Zurek1998, Schlosshauer2007Book}. In the entangling-speed-threshold theory, this question is answered by answering the question of where collapse occurs. After reviewing the dependence of entanglement on the tensor product structures, we will provide an answer to this question.

\begin{figure}[tb]
        \centering\includegraphics[width=7cm]{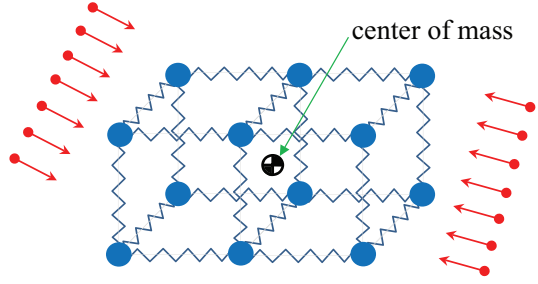}
        \caption{Tensor product structures for a system composed of a solid and environmental particles. TPS1, as given in Eq.~(1), decomposes the solid into individual atomic degrees of freedom, whereas TPS2, as given in Eq.~(2), decomposes the solid into the center-of-mass degree of freedom and phonon modes.}
\end{figure}

Given a suitably large Hilbert space, there are many legitimate ways to construct a tensor product structure to define the subsystems \cite{Zanardi2001, Zanardi2004}. Depending on how this construction is done, the amount of entanglement (and the entangling speed) varies. As an example, let us consider the situation illustrated in Fig.~1, where a solid consisting of $M$ atoms is interacting with many environmental particles. Among the many possible tensor product structures, we here investigate two representatives. The first is such that the solid is decomposed into individual atomic degrees of freedom, as follows: 
\begin{equation}
{\rm{TPS1}} \equiv \left[ {\mathop  \otimes \limits_{i = 1}^M \left( {{{\cal H}_{ix}} \otimes {{\cal H}_{iy}} \otimes {{\cal H}_{iz}}} \right)} \right] \otimes {{\cal H}_{\cal E}},
\end{equation}
where ${{\cal H}_{i\alpha }}$ $\left( {\alpha  = x,y,z} \right)$ is the ${i^{{\rm{th}}}}$ atomic Hilbert space and ${{\cal H}_{\cal E}}$ is the Hilbert space of the environmental degrees of freedom. In TPS1, ${{\cal H}_{i\alpha }}$ and the remainder can form a bipartite tensor product structure. In this bipartite cut, because ${{\cal H}_{i\alpha }}$ can interact only with neighboring particles, the entangling speed of ${{\cal H}_{i\alpha }}$ cannot be large. Now, we consider the second tensor product structure, in which the solid is decomposed into the center-of-mass degree of freedom and phonon modes, as follows:
\begin{equation}
{\rm{TPS2}} \equiv {{\cal H}_{CMx}} \otimes {{\cal H}_{CMy}} \otimes {{\cal H}_{CMz}} \otimes \left( {\mathop  \otimes \limits_{j = 1}^{3M - 3} {{\cal H}_j}} \right) \otimes {{\cal H}_{\cal E}},
\end{equation}
where ${{\cal H}_{CM\alpha }}$ $\left( {\alpha  = x,y,z} \right)$ is the Hilbert space of the center-of-mass degree of freedom and ${{\cal H}_j}$ is the Hilbert space of the ${j^{{\rm{th}}}}$ phonon mode. In TPS2, ${{\cal H}_{CM\alpha }}$ and the remainder can form a bipartite tensor product structure. In this bipartite cut, the entangling speed of  ${{\cal H}_{CM\alpha }}$ can be large because it interacts with many environmental particles simultaneously. In Section~VII, using a spin model, we will see that the entangling speed reaches a higher value as the number of environmental particles increases.

Now, we answer the question of where and when collapse occurs. In the example shown in Fig.~1, the entangling speed of ${{\cal H}_{CM\alpha }}$ in TPS2 would be larger than that of ${{\cal H}_{i\alpha }}$ in TPS1. Hence, if we postulate that collapse occurs when the entangling speed reaches a threshold, then the entangling speed of ${{\cal H}_{CM\alpha }}$ reaches this threshold first such that ${{\cal H}_{CM\alpha }}$ is where collapse occurs. In this way, ${{\cal H}_{CM\alpha }}$ appears to take on a state of reality, and TPS2, rather than TPS1, becomes the {\it{actual}} tensor product structure. Throughout this paper, we will call the degree of freedom in which collapse occurs the ``epicenter of collapse''.

The entangling-speed-threshold theory postulates that the epicenter of collapse is always a single degree of freedom, not a combined degree of freedom. By `single degree of freedom', we mean a non-decomposable Hilbert space. For example, ${{\cal H}_{ix}} \otimes {{\cal H}_{kx}}$ in TPS1 is a combined degree of freedom, and we do not consider it as a candidate to be the epicenter of collapse. By contrast, ${{\cal H}_{CM\alpha }}$ of TPS2 is a non-decomposable Hilbert space although all of the atomic coordinates participate in defining ${{\cal H}_{CM\alpha }}$, so it can be an epicenter of collapse. The reason we restrict the epicenter of collapse to be a single degree of freedom is that without this restriction, conflicts would arise in the theory. For example, imagine two systems ${{\cal H}_A}$ and ${{\cal H}_B}$ that are separated by a long distance, do not interact each other, and have their own environments with which they are interacting. In that case, there could be a situation in which neither ${{\cal H}_A}$ nor ${{\cal H}_B}$ reaches the entangling speed threshold, whereas ${{\cal H}_{A}} \otimes {{\cal H}_{B}}$ exceeds the threshold. Hence, if we allow ${{\cal H}_{A}} \otimes {{\cal H}_{B}}$ to be an epicenter of collapse, the quantum states of ${{\cal H}_A}$ and ${{\cal H}_B}$ would collapse as a result of the collapse of ${{\cal H}_{A}} \otimes {{\cal H}_{B}}$. This supposition is contradictory because it means that simple human consideration that regards the two systems as a combined one could cause a physical collapse. For this reason, we restrict the epicenter of collapse to only a single degree of freedom. The above discussions lead us to the first postulate of the theory.
 
\begin{itemize}
\item $\mathbf{Postulate~1}$: In any bipartite tensor product structure composed of a single degree of freedom and the remainder of the universe, when the entangling speed of the single degree of freedom reaches a threshold, the quantum state of that single degree of freedom collapses into a product state.
\end{itemize}

Whether the threshold is a universal constant or a system-dependent quantity should be judged through experiment. At this time, there is no theoretical reasoning that establishes the threshold as a universal constant, although we expect it to be. After successful experimental verification of the entangling-speed-threshold theory, the specific value(s) of the threshold could be measured in many different situations. 

A question arises regarding the implications of a negative entangling speed. Although the amount of entanglement is always positive, the entangling speed can be negative. If we interpret Postulate~1 as stating that a rapid growth in entanglement acts as a stress, forcing the system to collapse to release that stress, then we should conclude that a negative entangling speed will never suffice the collapse condition of Postulate~1 because a rapid decrease in entanglement would mean that the system is releasing stress. 

According to the entangling-speed-threshold theory, collapse occurs not because of the properties of a system itself but because of the interaction with its environment. Viewed in this way, our theory can be considered as an interaction-induced collapse theory. Without interaction, the entangling speed vanishes although entanglement survives. This is another reason why we chose the entangling speed as the collapse threshold rather than the amount of entanglement. The notion of the interaction-induced collapse is in stark contrast with the Ghirardi-Rimini-Weber theory \cite{Ghirardi1986} because their theory is a spontaneous collapse theory.

\subsection{How the collapse basis is determined}

The next question the theory should answer is how the collapse basis is determined. Although the term ``pointer basis'' has prevailed in the decoherence program \cite{Zurek1981}, we instead use the term ``collapse basis'' because our theory explicitly assumes objective collapse. In Postulate~1, the entangling speed plays the role of a stress that forces the single degree of freedom to collapse. Hence, it is natural to assume that collapse processes are stress-releasing processes in which the collapse basis is determined so as to eliminate the stress. Before we see how this process is achieved, let us first review the characteristics of a product state. A product state, by definition, has no entanglement. Furthermore, the entangling speed is also zero for a product state. This fact is because von Neumann entropy (the measure of entanglement that we use throughout this paper) is a differentiable function of time and has a minimum (zero) at a product state (note that the time derivative of a differentiable function at a minimum point is always zero). Hence, as long as we assume that collapse occurs into a product state, both the entanglement and the entangling speed vanish automatically, regardless of the particular choice of the collapse basis. If we require robustness against entanglement for adequacy as a collapse basis, it is natural to require the next-order time derivative of entanglement, i.e., the entangling acceleration, to vanish or at least to be minimized. In fact, the entangling acceleration is a basis-dependent quantity. Hence, the requirement of a minimum entangling acceleration can serve to select a particular basis. Consequently, we postulates that the collapse basis is determined such that the entangling acceleration is minimized immediately after the collapse. 

Let us make a mathematical statement about how the collapse basis is determined. Consider a bipartite tensor product structure composed of a single degree of freedom ${\cal A}$ whose Hilbert space dimension is ${d_{\cal A}}$, and an environment ${\cal E}$. Suppose that at time ${t_C}$, the quantum state of the entire system is $\left| {\Psi ({t_C})} \right\rangle $ and the entangling speed of ${\cal A}$ reaches the threshold. This supposition means that at ${t_C}$, ${\cal A}$ fulfills the collapse condition of Postulate~1 and is about to collapse into a product state. We can represent $\left| {\Psi ({t_C})} \right\rangle $ using the supposed collapse basis $\left\{ {\left| {{{\cal A}_i}} \right\rangle } \right\}$ of ${\cal A}$ as follows:
\begin{equation}
\left| {\Psi ({t_C})} \right\rangle  = \sum\limits_{i = 1}^{{d_{\cal A}}} {{c_i}\left| {{{\cal A}_i}} \right\rangle \left| {{{\cal E}_i}} \right\rangle } ,
\end{equation}
where $\left\{ {\left| {{{\cal E}_i}} \right\rangle } \right\}$ is not necessarily an orthogonal set, but the relative states corresponding to the orthonormal basis $\left\{ {\left| {{{\cal A}_i}} \right\rangle } \right\}$. Hence, Eq.~(3) is not a Schmidt decomposition. $\left| {\Psi ({t_C})} \right\rangle $ is assumed to collapse into one of the product states $\left| {{{\cal A}_i}} \right\rangle \left| {{{\cal E}_i}} \right\rangle $. Then, immediately after the collapse, the product state begins to evolve with the unitary evolution governed by the total Hamiltonian $\hat H$. The amount of entanglement ${\epsilon_i}$ (the von Neumann entropy of ${\cal A}$) and its time derivatives ${\dot \epsilon_i}$ and ${\ddot \epsilon_i}$ should vary deterministically under this unitary evolution. We focus only on the entangling acceleration ${\ddot \epsilon_i}$ at the instant immediately after the collapse because both ${\epsilon_i}$ and ${\dot \epsilon_i}$ are zero at that instant. Now, we define the ensemble-averaged entangling acceleration $\bar{\ddot \epsilon}$ as
\begin{equation}
{\bar{\ddot \epsilon}}  \equiv \sum\limits_i^{{d_{\cal A}}} {{{\left| {{c_i}} \right|}^2}{{\ddot \epsilon}_i}} .
\end{equation}
We postulate that the collapse basis $\left\{ {\left| {{{\cal A}_i}} \right\rangle } \right\}$ is a basis that minimizes $\bar{\ddot \epsilon}$. We will see later that there exists a unique $\left\{ {\left| {{{\cal A}_i}} \right\rangle } \right\}$ that causes $\bar{\ddot \epsilon}$ to be near zero when the environment consists of many degrees of freedom. In this way, the collapse basis is determined, and $\left| {\Psi ({t_C})} \right\rangle $ collapses into $\left| {{{\cal A}_i}} \right\rangle \left| {{{\cal E}_i}} \right\rangle $ with probability ${\left| {{c_i}} \right|^2}$, which is Born's rule. The above discussions lead to the second postulate of the theory.

\begin{itemize}
\item $\mathbf{Postulate~2}$: The collapse basis is determined such that, immediately after collapse, the ensemble-averaged entangling acceleration is minimized.
\end{itemize}

In a realistic situation in which the environment ${\cal E}$ consists of many degrees of freedom, we can find a more explicit formula for determining the collapse basis. Immediately after collapse, each collapse state $\left| {{{\cal A}_i}} \right\rangle \left| {{{\cal E}_i}} \right\rangle $ must be the state that is most robust against entanglement formation. Let us consider the infinitesimal time evolution of a collapse state $\left| {{{\cal A}_i}} \right\rangle \left| {{{\cal E}_i}} \right\rangle $ (using units of $\hbar  = 1$). 
\begin{gather}
{e^{ - i\hat H\Delta t}}\left| {{{\cal A}_i}} \right\rangle \left| {{{\cal E}_i}} \right\rangle \simeq \left( {1 - i\hat H\Delta t} \right)\left| {{{\cal A}_i}} \right\rangle \left| {{{\cal E}_i}} \right\rangle \nonumber
\\
= \left| {{{\cal A}_i}} \right\rangle \left| {{{\cal E}_i}} \right\rangle  - i\Delta t\sum\limits_{j,k} {\langle {{\cal A}_j}|\langle {\cal E}_k^{(i)}|\hat H\left| {{{\cal E}_i}} \right\rangle \left| {{{\cal A}_i}} \right\rangle  \cdot \left| {{{\cal A}_j}} \right\rangle \left| {{\cal E}_k^{(i)}} \right\rangle } . 
\end{gather}
In the second line of the above equation, we used $\sum\limits_{j = 1}^{{d_{\cal A}}} {\sum\limits_{k = 1}^{{d_{\cal E}}} {\left| {{\cal E}_k^{(i)}} \right\rangle \left| {{{\cal A}_j}} \right\rangle \langle {{\cal A}_j}|\langle {\cal E}_k^{(i)}|}  = \hat I} $, where ${d_{\cal E}}$ is the Hilbert space dimension of the environment and $\left\{ {\left| {{\cal E}_k^{(i)}} \right\rangle } \right\}$ is any orthonormal basis of ${\cal E}$ that contains $\left| {{{\cal E}_i}} \right\rangle $ (note that $\left\{ {\left| {{{\cal E}_i}} \right\rangle } \right\}$ in Eq.~(3) and $\left\{ {\left| {{\cal E}_k^{(i)}} \right\rangle } \right\}$ here are different). For the evolved state to have minimal entanglement, $\left\langle {{{\cal A}_{j \ne i}}} \right|\left\langle {{\cal E}_{k \ne i}^{(i)}} \right|\hat H\left| {{{\cal E}_i}} \right\rangle \left| {{{\cal A}_i}} \right\rangle $ in Eq.~(5) should be minimized. The simplest way to achieve this minimization is that  ${\left| {{{\cal A}_i}} \right\rangle }$ becomes an eigenstate of $\left\langle {{\cal E}_{k \ne i}^{(i)}} \right|\hat H\left| {{{\cal E}_i}} \right\rangle $ so as that $\left\langle {{{\cal A}_{j \ne i}}} \right|\left\langle {{\cal E}_{k \ne i}^{(i)}} \right|\hat H\left| {{{\cal E}_i}} \right\rangle \left| {{{\cal A}_i}} \right\rangle $ vanishes. To be more specific, let us express the total Hamiltonian $\hat H$ in the most general form as follows: 
\begin{equation}
\hat H = {\hat H_{\cal A}} \otimes {\hat I_{\cal E}} + \sum\limits_\alpha  {{{\hat {\cal A}}_\alpha } \otimes {{\hat {\cal E}}_\alpha }}  + {\hat I_{\cal A}} \otimes {\hat H_{\cal E}},
\end{equation}
where the second term is the interaction Hamiltonian ${\hat H_{{\rm{int}}}}$ which is written in the form of a diagonal decomposition of the ${\cal A}$- and ${\cal E}$-operators ${\hat {\cal A}_\alpha }$ and ${\hat {\cal E}_\alpha }$, respectively. With this form of $\hat H$, we obtain
\begin{equation}
\left\langle {{\cal E}_{k \ne i}^{(i)}} \right|\hat H\left| {{{\cal E}_i}} \right\rangle  = \left\langle {{\cal E}_{k \ne i}^{(i)}} \right|\left. {{{\cal E}_i}} \right\rangle {\hat H_{\cal A}} + \sum\limits_\alpha  {\left\langle {{\cal E}_{k \ne i}^{(i)}} \right|{{\hat {\cal E}}_\alpha }\left| {{{\cal E}_i}} \right\rangle {{\hat {\cal A}}_\alpha }}  + \left\langle {{\cal E}_{k \ne i}^{(i)}} \right|{\hat H_{\cal E}}\left| {{{\cal E}_i}} \right\rangle {\hat I_{\cal A}}.
\end{equation}
The first term vanishes because $\left\langle {{\cal E}_{k \ne i}^{(i)}} \right|\left. {{{\cal E}_i}} \right\rangle  = 0$. Moreover, because the last term is a scaled identity operator, if $\left| {{{\cal A}_i}} \right\rangle $ is an eigenstate of the second term, then $\left\langle {{{\cal A}_{j \ne i}}} \right|\left\langle {{\cal E}_{k \ne i}^{(i)}} \right|\hat H\left| {{{\cal E}_i}} \right\rangle \left| {{{\cal A}_i}} \right\rangle $ will be zero. 

Now, we use the fact that the environment ${\cal E}$ consists of many degrees of freedom. In general, a Hamiltonian with many degrees of freedom is highly degenerate \cite{Nikolic2015}. This means that the $\left\langle {{\cal E}_{k \ne i}^{(i)}} \right|{\hat {\cal E}_\alpha }\left| {{{\cal E}_i}} \right\rangle $s are nearly identical regardless of the particular choice of $\left\langle {{\cal E}_{k \ne i}^{(i)}} \right|$ and $\left| {{{\cal E}_i}} \right\rangle $. Therefore, we can safely approximate  
\begin{equation}
\left\langle {{\cal E}_{k \ne i}^{(i)}} \right|{\hat {\cal E}_\alpha }\left| {{{\cal E}_i}} \right\rangle  \approx \left\langle {{{\cal E}_0}} \right|{\hat {\cal E}_\alpha }\left| {{{\cal E}_0}} \right\rangle ,
\end{equation}
where $\left| {{{\cal E}_0}} \right\rangle $ is a typical environmental state. Now, we define a ``collapse operator'' $\hat C$ such that
\begin{equation}
\hat C \equiv \left\langle {{{\cal E}_0}} \right|{\hat H_{{\rm{int}}}}\left| {{{\cal E}_0}} \right\rangle  = \sum\limits_\alpha  {\left\langle {{{\cal E}_0}} \right|{{\hat {\cal E}}_\alpha }\left| {{{\cal E}_0}} \right\rangle {{\hat {\cal A}}_\alpha }}, 
\end{equation}
whose eigenbasis is effectively the collapse basis $\left\{ {\left| {{{\cal A}_i}} \right\rangle } \right\}$ that satisfies Postulate~2. Because $\hat C$ is Hermitian, $\left\{ {\left| {{{\cal A}_i}} \right\rangle } \right\}$ is guaranteed to be an orthonormal set. 

The determination of the collapse basis is unrelated to the self-Hamiltonians of ${\cal A}$ and ${\cal E}$. In Eq.~(9), the collapse operator $\hat C$ involves only the interaction Hamiltonian ${\hat H_{{\rm{int}}}}$. Nevertheless, the self-Hamiltonians play a crucial role in the unitary time evolution that happens during the time interval between collapses. Through this unitary evolution, the next collapse instant is determined. 

In determining the collapse basis, it is worth noting the similarities and differences between the decoherence program and the entangling-speed-threshold theory. The main similarity is that both theories require robustness against entanglement for the pointer (collapse) basis. However, with regard to this robustness requirement, the ``predictability sieve'' strategy of the decoherence program demands robustness for a long time; therefore, the strategy relies on a time-integrated expression \cite{Busse2010Thesis}. By contrast, the entangling-speed-threshold theory requires robustness only at the instant immediately after the collapse; therefore, the collapse basis is determined by a time-local condition.

\section{What are the systems?}

If the universe were not resolved into individual subsystems, the measurement problem would disappear because there would be no need for ``collapse'' in the universe with no systems \cite{Zurek1998, Schlosshauer2007Book}. The measurement problem arises only because a part of the universe appears to have an individual reality. In Subsection~II.A, we explained how a large Hilbert space is divided into subsystems (determining the {\it{actual}} tensor product structure). In this section, we further discuss about the meaning of systems.

We propose a definition regarding the identity of systems: a system is a degree of freedom that collapses. For a system to be an {\it{actual}} system, we believe that the system should be able to be measured by the remainder of the universe, and collapse is a necessary capability for measurement. From this viewpoint, two kinds of systems exist because there are two kinds of collapse. Systems of the first kind are epicenters of collapse, which satisfy the collapse condition of Postulate~1. Most macroscopic systems, such as the center-of-mass degree of freedom of a solid, are systems of this kind. In addition to epicentral collapse, there is another type of collapse that occurs when certain degrees of freedom entangle with an epicenter of collapse. When an epicenter of collapse collapses, these entangled degrees of freedom also collapse. We call a collapse of this type an ``affected collapse'', which defines systems of the second kind. Most microscopic systems, such as atoms or photons, are systems of the second kind, whose reality arises with the aid of a macroscopic epicentral collapse system. 

Unlike the Ghirardi-Rimini-Weber theory \cite{Ghirardi1986}, our theory predicts that even a massive macroscopic object can exhibit quantum behavior as long as it is well isolated from its environment. Progress in technology has already yielded several indications in this direction, e.g., in experiments using a $\rm{C}_{70}$ molecule \cite{Bbrezger2002} or a mechanical resonator \cite{OConnell2010}. However, even a single environmental particle can ruin the quantum coherence of a macroscopic system if they interact and become entangled. After entanglement is formed, if the environmental particle encounters another epicenter of collapse, it will undergo affected collapse, and the macroscopic system will also undergo affected collapse and lose its quantum coherence. Because macroscopic systems are highly prone to interaction with other objects, it is difficult for a macroscopic system to maintain its quantum coherence.

\section{Quantum-to-classical transition}

The entangling-speed-threshold theory can explain how macroscopic objects behave classically. Whether a system is quantum or classical depends on how frequently they collapse. If the time interval between collapses is long, the system can behave quantum mechanically during that time interval. Otherwise, if the time interval is short, quantum coherence will disappear so rapidly that the system will behave classically. Based on this consideration, most epicentral collapse systems will behave classically, whereas affected collapse systems can exhibit quantum behavior.

\begin{figure}[tb]
        \centering\includegraphics[width=7cm]{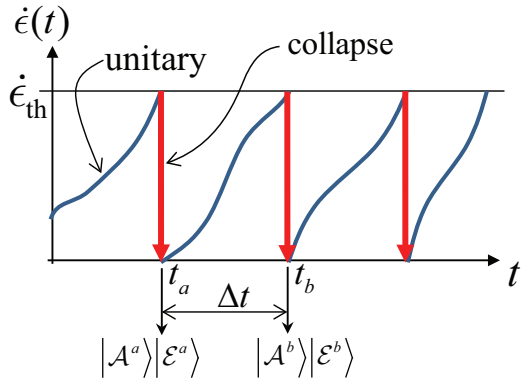}
        \caption{Emergence of deterministic classical dynamics from indeterministic quantum collapse. $\dot \epsilon$ is the entangling speed of a macroscopic system and ${\dot \epsilon_{{\rm{th}}}}$ is the threshold at which collapse occurs. Macroscopic epicentral collapse systems collapse very frequently ($\Delta t$ may be shorter than ${10^{ - 18}}$~sec). Due to this nearly continuous collapse, after collapsing into $\left| {{{\cal A}^a}} \right\rangle \left| {{{\cal E}^a}} \right\rangle $ at ${t_a}$, the next collapse state $\left| {{{\cal A}^b}} \right\rangle \left| {{{\cal E}^b}} \right\rangle $ arises with a high Born's probability near 1. In this way, the slight state change $\left| {{{\cal A}^a}} \right\rangle \buildrel {\Delta t} \over
 \longrightarrow \left| {{{\cal A}^b}} \right\rangle $ appears to be governed by deterministic classical dynamics.}
\end{figure}

Here, we focus on deriving deterministic classical dynamics, which seems irreconcilable with indeterministic quantum collapse. The key to reconciling these two opposing concepts, determinism and indeterminism, is the fact that a certain collapse state happens with a high Born's probability near 1. A macroscopic system, e.g., the center of mass of a solid, simultaneously interacts with many environmental particles (imagine Avogadro's number of air molecules, photons, or cosmic rays). Its entangling speed can increase very rapidly after it collapses into a product state. Therefore, a macroscopic system should collapse nearly continuously. Figure~2 illustrates this situation. Suppose that at ${t_a}$, the macroscopic system satisfies the collapse condition and collapses into a product state $\left| {{{\cal A}^a}} \right\rangle \left| {{{\cal E}^a}} \right\rangle $. Then, $\left| {{{\cal A}^a}} \right\rangle \left| {{{\cal E}^a}} \right\rangle $ evolves unitarily with the total Hamiltonian and will satisfy the collapse condition again at ${t_b}$. The collapse basis is determined as the eigenbasis of the collapse operator $\hat C$ of Eq.~(9). If $\Delta t$ is very small, $\hat C$ at ${t_b}$ should be close to $\hat C$ at ${t_a}$, meaning that the collapse bases are almost the same at ${t_a}$ and ${t_b}$. Moreover, the unitarily evolved state of ${\cal A}$ should be close to $\left| {{{\cal A}^a}} \right\rangle $ although the unitary evolution would make the state of ${\cal A}$ slightly mixed. Consequently, at ${t_b}$, there should exist a certain collapse state $\left| {{{\cal A}^b}} \right\rangle \left| {{{\cal E}^b}} \right\rangle $ that has a high Born's probability near 1. Thus, $\left| {{{\cal A}^b}} \right\rangle \left| {{{\cal E}^b}} \right\rangle $ is chosen almost deterministically. If we focus on the state change $\left| {{{\cal A}^a}} \right\rangle \buildrel {\Delta t} \over
 \longrightarrow \left| {{{\cal A}^b}} \right\rangle $, this slight change will appear as if it is governed by deterministic classical dynamics, in which $\left| {{{\cal A}^a}} \right\rangle $ and $\left| {{{\cal A}^b}} \right\rangle $ are classically preferred states. 

The above discussion is reminiscent of the quantum Zeno effect in which the state of a system freezes by frequent measurements \cite{Misra1977}. Indeed, many classical systems appear to keep their states without evolution. For example, a pencil is stationary on the desk, and a photon detector maintains the ready state until it detects a photon. We can regard these examples as special cases of Fig.~2. 

We predict that for a classical system, the time interval $\Delta t$ between collapses may be shorter than ${10^{ - 18}}$~sec. With this extremely short time interval, our classical world appears to exhibit continuous dynamics. We also argue that the ultimate resolution of a clock is limited by its $\Delta t$, which depends on how strongly the clock interacts with the environment.

\section{Measurement}

Measurement is a process that involves three parties: the measured system ${\cal S}$, the measuring apparatus ${\cal A}$, and the environment ${\cal E}$. To illustrate this process more concretely, we consider the example presented in Fig.~3, where a photon detector ${\cal A}$ detects a photon ${\cal S}$. Regardless of ${\cal S}$, ${\cal A}$ continuously satisfies the collapse condition through its interaction with ${\cal E}$. We assume that the collapse basis is determined to be $\left\{ {\left| {{\rm{unclick}}} \right\rangle ,\left| {{\rm{click}}} \right\rangle } \right\}$. Before ${\cal S}$ interacts with ${\cal A}$, ${\cal A}$ is continuously collapsing into $\left| {{\rm{unclick}}} \right\rangle $ due to the quantum Zeno effect. Now, suppose that the interaction between ${\cal S}$ and ${\cal A}$ causes an evolution such that 
\begin{equation}
\left| 1 \right\rangle \left| {{\rm{unclick}}} \right\rangle \buildrel {\Delta t} \over
 \longrightarrow \sqrt {1 - {{\left| {{c_{{\rm{click}}}}} \right|}^2}} \left| 1 \right\rangle \left| {{\rm{unclick}}} \right\rangle  + {c_{{\rm{click}}}}\left| 0 \right\rangle \left| {{\rm{click}}} \right\rangle ,
\end{equation}
where $\left| 1 \right\rangle $ and $\left| 0 \right\rangle $ are the single-photon state and the vacuum state, respectively, and $\Delta t$ is the time interval between collapses . During $\Delta t$, the photon field and the detector are slightly entangled, and at the next collapse, there is a probability ${p_{{\rm{click}}}} = {\left| {{c_{{\rm{click}}}}} \right|^2}$ for the detector to collapse into $\left| {{\rm{click}}} \right\rangle $. A sensible detector should be designed such that ${p_{{\rm{click}}}}$ is proportional to the intensity of the photon field. Hence, as long as $\Delta t$ is sufficiently short compared with the total span of the photon field, ${p_{{\rm{click}}}}$ should be proportional to the intensity of the photon field at the very instant of collapse. Therefore, the photon intensity profile of Fig.~3(a) and the distribution of ${p_{{\rm{click}}}}$ in 3(b) are identical (the reverse in shape is because they are spatial and temporal distributions, respectively). When the detector collapses into $\left| {{\rm{click}}} \right\rangle $, the photon field also collapses into $\left| 0 \right\rangle $ via affected collapse. Once $\left| 0 \right\rangle \left| {{\rm{click}}} \right\rangle $ is established, this state remains for a long time because of the quantum Zeno effect and because of the lack of any further photons. A practical detector usually has a recovery circuit that returns the state of the detector to $\left| {{\rm{unclick}}} \right\rangle $ to be prepared for the next detection; the time needed for this process is called the dead time. 

\begin{figure}[tb]
        \centering\includegraphics[width=7cm]{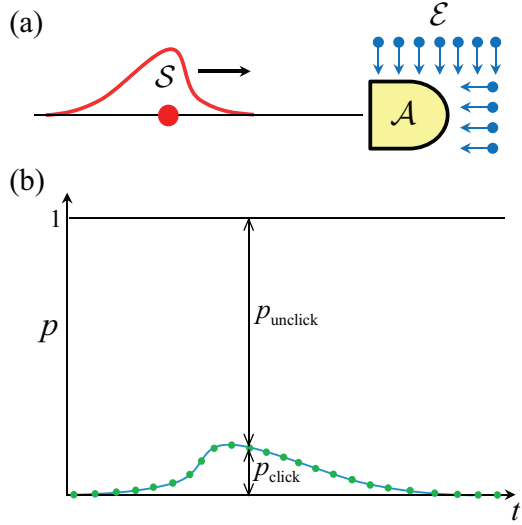}
        \caption{Photon detection measurement. (a) Measurement setup. A photon ${\cal S}$ is entering a detector ${\cal A}$ that is interacting with an environment ${\cal E}$. (b) Probability distributions of ${p_{{\rm{click}}}}$ and ${p_{{\rm{unclick}}}}$ at each collapse. The filled green circles indicate the instants of collapse. The distribution of ${p_{{\rm{click}}}}$ is proportional to the intensity profile of the photon field.}
\end{figure}

In our collapse theory, observables (or, more generally, measurement operators) are derived from how ${\cal S}{\cal A}{\cal E}$ interact. ${\cal A}$ plays the role of an epicenter of collapse that is continuously collapsing through its interaction with others. Due to the myriad of environmental particles, the entangling speed of ${\cal A}$ is mainly determined by the ${\cal A}{\cal E}$ interaction, so ${\cal S}$ is negligible in determining the collapse instants and the collapse basis. Suppose that the collapse basis (i.e., the possible pointer states of ${\cal A}$) has been determined to be $\left\{ {\left| {{{\cal A}_m}} \right\rangle } \right\}$, and also suppose that before the measurement, ${\cal A}$ is continuously collapsing into the ready state $\left| {{{\cal A}_r}} \right\rangle  \in \left\{ {\left| {{{\cal A}_m}} \right\rangle } \right\}$ due to the quantum Zeno effect. Before ${\cal S}$ enters ${\cal A}$, the initial state of ${\cal S}{\cal A}$ is a product state $\left| {{{\cal S}_0}} \right\rangle \left| {{{\cal A}_r}} \right\rangle $, where $\left| {{{\cal S}_0}} \right\rangle $ is an arbitrary initial state of ${\cal S}$. Now, ${\cal S}$ begins to interact with ${\cal A}$, and they begin to entangle through unitary evolution. Through the continuous collapse, ${\cal A}$ will predominantly collapse into $\left| {{{\cal A}_r}} \right\rangle $ at each collapse. However, a tiny probability of collapsing into $\left| {{{\cal A}_{m \ne r}}} \right\rangle $ accumulates up through the end of the measurement. Mostly, we are interested in the final accumulated probability of $\left| {{{\cal A}_{m \ne r}}} \right\rangle $. The final accumulated probability can be obtained by considering the case in which no collapse occurs and unitary evolution continues through the end of the measurement. This hypothetical unitary evolution of ${\cal S}{\cal A}$ up through the end of the measurement can be written as 
\begin{equation}
\left| {{{\cal S}_0}} \right\rangle \left| {{{\cal A}_r}} \right\rangle  \to \sum\limits_m {{c_m}\left| {{{\cal S}_m}} \right\rangle \left| {{{\cal A}_m}} \right\rangle } ,
\end{equation}
where $\left\{ {\left| {{{\cal S}_m}} \right\rangle } \right\}$ is not necessarily an orthogonal set but the relative states corresponding to the predetermined orthonormal collapse basis $\left\{ {\left| {{{\cal A}_m}} \right\rangle } \right\}$. The final measurement outcome will be a particular state ${\left| {{{\cal S}_m}} \right\rangle \left| {{{\cal A}_m}} \right\rangle }$ with probability ${p_m} = {\left| {{c_m}} \right|^2}$. If we observe a measurement outcome $\left| {{{\cal A}_{m}}} \right\rangle $, we infer that ${\cal S}$ is measured resulting in wavefunction collapse into $\left| {{{\cal S}_m}} \right\rangle$. The measurement operators $\left\{ {{M_m}} \right\}$ of ${\cal S}$ are derived such that 
\begin{equation}
{M_m}\left| {{{\cal S}_0}} \right\rangle  = {c_m}\left| {{{\cal S}_m}} \right\rangle .
\end{equation}
It can be easily shown that ${M_m}$ satisfies the two necessary conditions required for measurement operators \cite{Nielsen2000Book}: ${p_m} = \left\langle {{{\cal S}_0}} \right|M_m^\dag {M_m}\left| {{{\cal S}_0}} \right\rangle $ and $\sum\limits_m {M_m^\dag {M_m} = I} $. If, in Eq.~(12), ${M_m} = \left| {{{\cal S}_m}} \right\rangle \left\langle {{{\cal S}_m}} \right|$ is satisfied, then the measurement becomes a projective measurement. In summary, the collapse basis of ${\cal A}$ is determined by how ${\cal A}{\cal E}$ interact, and the observables (measurement operators) of ${\cal S}$ are derived by how ${\cal S}{\cal A}$ interact and by the collapse basis. 

Suppose that we want to detect a particular target state $\left| {{{\cal S}_m}} \right\rangle $ from the input state $\left| {{{\cal S}_0}} \right\rangle $. The probability of detecting $\left| {{{\cal S}_m}} \right\rangle $ accumulates as ${\cal S}$ and ${\cal A}$ interact. For an ideal detector, the final accumulated probability should be identical to ${\left| {\left\langle {{{\cal S}_m}} \right.\left| {{{\cal S}_0}} \right\rangle } \right|^2}$, which is difficult to achieve.

Among many instants of collapse, one particular instant becomes the instant of measurement. This is because measurement is achieved when a measuring apparatus ${\cal A}$ collapses from the ready state $\left| {{{\cal A}_r}} \right\rangle $ to another discernible collapse state $\left| {{{\cal A}_{m \ne r}}} \right\rangle $. The collapse from $\left| {{{\cal A}_r}} \right\rangle $ to $\left| {{{\cal A}_{m \ne r}}} \right\rangle $ is called a quantum jump, which corresponds to an escape from the quantum Zeno effect. Hence, the instant of measurement is determined probabilistically, whereas the instants of collapse are determined deterministically by Postulate~1. We previously observed this phenomenon in the example of Fig.~3, in which the instant of photon detection arises probabilistically. 

The Copenhagen interpretation requires the existence of a classical measuring apparatus which is regarded as an entity outside the quantum mechanical framework. In contrast, the entangling-speed-threshold theory treats all entities in the universe on the same theoretical framework. The only difference between quantum objects and classical objects is whether they satisfy Postulate~1 to serve as an epicenter of collapse. A measuring apparatus is a specially designed epicentral collapse system that is capable of performing a quantum jump from the ready state to another discernible collapse state when triggered by the introduction of a tiny measured system.

\section{Energy conservation}

If quantum collapses are realistic objective processes, then energy should be conserved before and after collapse. Ref. \cite{Nikolic2015} presented a proof of energy conservation for nonunitary processes in which environment consists of many degrees of freedom. Here, we provide a different proof that the entangling-speed-threshold theory guarantees energy conservation to a high accuracy, although not exactly. Suppose that an epicenter of collapse ${\cal A}$ fulfills the collapse condition of Postulate~1 by interacting with its environment ${\cal E}$. Let the state of the entire system, immediately before the collapse, be represented by Eq.~(3). Then, the energy expectation value before the collapse is
\begin{equation}
{E_{{\rm{before}}}} = \sum\limits_{i,j}^{{d_{\cal A}}} {c_j^ * {c_i}\left\langle {{{\cal A}_j}} \right|\left\langle {{{\cal E}_j}} \right|\hat H\left| {{{\cal E}_i}} \right\rangle \left| {{{\cal A}_i}} \right\rangle } ,
\end{equation}
where $\hat H$ is the total Hamiltonian given in the form of Eq.~(6). Because, after the collapse, the state of the entire system is one of the product states $\left\{ {\left| {{{\cal A}_i}} \right\rangle \left| {{{\cal E}_i}} \right\rangle } \right\}$, the ensemble-averaged energy expectation value immediately after the collapse will be
\begin{equation}
{E_{{\rm{after}}}} = \sum\limits_i^{{d_{\cal A}}} {{{\left| {{c_i}} \right|}^2}\left\langle {{{\cal A}_i}} \right|\left\langle {{{\cal E}_i}} \right|\hat H\left| {{{\cal E}_i}} \right\rangle \left| {{{\cal A}_i}} \right\rangle } .
\end{equation}
Hence, the energy difference before and after the collapse is 
\begin{equation}
\Delta E = \sum\limits_{i \ne j}^{{d_{\cal A}}} {c_j^ * {c_i}\left\langle {{{\cal A}_j}} \right|\left\langle {{{\cal E}_j}} \right|\hat H\left| {{{\cal E}_i}} \right\rangle \left| {{{\cal A}_i}} \right\rangle } .
\end{equation}
Therefore, if $\left\langle {{{\cal A}_{j \ne i}}} \right|\left\langle {{{\cal E}_{j \ne i}}} \right|\hat H\left| {{{\cal E}_i}} \right\rangle \left| {{{\cal A}_i}} \right\rangle $ vanishes, $\Delta E$ will be zero. 

We now prove that $\left\langle {{{\cal A}_{j \ne i}}} \right|\left\langle {{{\cal E}_{j \ne i}}} \right|\hat H\left| {{{\cal E}_i}} \right\rangle \left| {{{\cal A}_i}} \right\rangle $ effectively vanishes if the environment consists of many degrees of freedom. In Section~II, the existence of many environmental degrees of freedom led us to the conclusion that $\left\{ {\left| {{{\cal A}_i}} \right\rangle } \right\}$ should be the eigenbasis of the collapse operator $\hat C$ of Eq.~(9). The same logic applies here. $\left\langle {{{\cal E}_{j \ne i}}} \right|\hat H\left| {{{\cal E}_i}} \right\rangle $ is nearly identical to Eq.~(7). Suppose that we replace $\left\langle {{\cal E}_{k \ne i}^{(i)}} \right|$ in Eq.~(7) with $\left\langle {{{\cal E}_{j \ne i}}} \right|$. Then, in the first term of Eq.~(7), $\left\langle {{{\cal E}_{j \ne i}}} \right|\left. {{{\cal E}_i}} \right\rangle $ would not necessarily be zero. However, $\left\langle {{{\cal E}_{j \ne i}}} \right|\left. {{{\cal E}_i}} \right\rangle {\hat H_{\cal A}}$ would be negligible compared with the collapse operator $\hat C$ of Eq.~(9). This result is because $\left\langle {{{\cal E}_{j \ne i}}} \right|\left. {{{\cal E}_i}} \right\rangle $ is at most 1, whereas $\left\langle {{{\cal E}_{j \ne i}}} \right|{\hat {\cal E}_\alpha }\left| {{{\cal E}_i}} \right\rangle $ is typically enormous when there are many environmental degrees of freedom. Hence, by the same reasoning through which we arrived at Eq.~(8) and (9), we find that $\left\langle {{{\cal E}_{j \ne i}}} \right|\hat H\left| {{{\cal E}_i}} \right\rangle  \approx \hat C$. Thus, $\left\langle {{{\cal A}_{j \ne i}}} \right|\left\langle {{{\cal E}_{j \ne i}}} \right|\hat H\left| {{{\cal E}_i}} \right\rangle \left| {{{\cal A}_i}} \right\rangle $ in Eq.~(15) effectively vanishes if $\left\{ {\left| {{{\cal A}_i}} \right\rangle } \right\}$ is the eigenbasis of the collapse operator $\hat C$. In other words, the collapse basis, which is the eigenbasis of $\hat C$, automatically satisfies energy conservation very accurately. In the next section, we will analyze a spin model and observe that $\Delta E$ becomes negligible as the number of environmental particles increases. 

The above argument guarantees energy conservation in only ensemble level because $E_{\rm{after}}$ of Eq.~(14) is the ensemble-averaged energy after collapse. Now, we prove that energy is well conserved even in a single collapse event if collapse occurs very frequently. In Section~IV, we observed that a certain collapse state arises with a high Born's probability near 1 bringing about deterministic classical dynamics. This deterministic collapse means that, in Eq.~(14), a certain ${\left| {{c_i}} \right|^2}$ dominates, and other probabilities are negligible. In other words, the ensemble-averaged energy is nearly identical to the energy of the single dominant outcome. Thus, when collapse brings about continuous and deterministic classical dynamics, energy is effectively conserved even in a single collapse event.

\section{Application to a spin model}

We now apply the entangling-speed-threshold theory to a simple spin model in which a spin-one-half system simultaneously interacts with many environmental spins. The purpose of this section is to give the reader a sense of the theory through a specific example. We will focus on the role of the number of environmental particles. We will see that as the number of environmental particles increases, (i) the entangling speed of the system increases and (ii) energy is more accurately conserved. 

We consider the situation in which ${\cal A}$, a spin-one-half system, interacts with an environment ${\cal E}$ consisting of $N$ spins. We assume that the total Hamiltonian is given by 
\begin{equation}
\hat H = {\hat \sigma _{x,{\cal A}}} + {\hat \sigma _{z,{\cal A}}} \otimes \sum\limits_{k = 1}^N {{{\hat \sigma }_{z,k}}}  + \sum\limits_{k = 1}^N {{{\hat \sigma }_{x,k}}} ,
\end{equation}
where the first and third terms are the self-Hamiltonians of ${\cal A}$ and ${\cal E}$, respectively, and the second term is the interaction Hamiltonian. We set the initial state to ${\left|  +  \right\rangle _{\cal A}} \otimes \left|  +  \right\rangle _{\cal E}^{ \otimes N}$ (where $\left|  +  \right\rangle  = {2^{ - 1/2}}\left( {\left| 0 \right\rangle  + \left| 1 \right\rangle } \right)$), and we numerically simulated the time evolution of the state. As Fig.~4(a) and (b) show, the amount of entanglement $\epsilon (t)$ (the von Neumann entropy of ${\cal A}$) and the entangling speed $\dot \epsilon (t)$ vary with time. As $N$ increases, $\dot \epsilon (t)$ rises more rapidly and reaches a higher value. The same tendency is observed even when the initial environmental state is random. Thus, with larger $N$, ${\cal A}$ is more likely to meet the collapse condition of Postulate~1. 

\begin{figure}[tb]
        \centering\includegraphics[width=9cm]{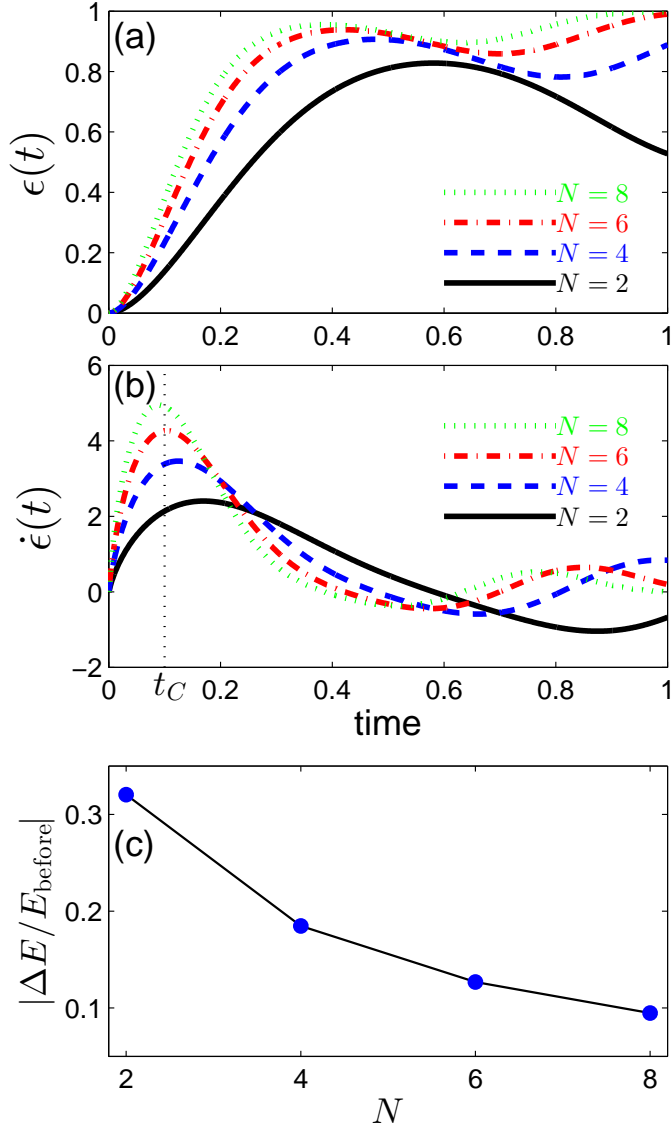}
        \caption{Numerical simulation of the spin model of Eq.~(16), with varying the number of environmental spins $N$. The initial state was set to ${\left|  +  \right\rangle _{\cal A}} \otimes \left|  +  \right\rangle _{\cal E}^{ \otimes N}$. (a) The amount of entanglement ${\epsilon}(t)$ (von-Neumann entropy) of ${\cal A}$. (b) The entangling speed ${\dot \epsilon}(t)$. As $N$ increases, ${\dot \epsilon}$ increases rapidly and gets to a higher value. (c) The energy deviation before and after collapse, which is assumed to occur at ${t_C}$, as marked in (b). As $N$ increases, energy is better conserved.}
\end{figure}

To investigate the issue of energy conservation, we chose a particular time ${t_C}$ as an instant of collapse, as marked in Fig.~4(b). We took the states at ${t_C}$ to be $\left| {\Psi ({t_C})} \right\rangle $ of Eq.~(3), i.e., the state immediately before the collapse. We then selected the collapse basis $\left\{ {\left| {{{\cal A}_i}} \right\rangle } \right\}$ that minimizes $\bar{\ddot \epsilon} $ in Eq.~(4) by numerically scanning the entire Hilbert space of ${\cal A}$. After selecting the collapse basis, we calculated the energy difference $\Delta E$ using Eq.~(15). Figure~4(c) plots $\left| {\Delta E/{E_{{\rm{before}}}}} \right|$ as a function of $N$. As $N$ increases, energy tends to be more accurately conserved. The entangling-speed-threshold theory guarantees energy conservation to a high accuracy as long as ${\cal A}$ is interacting with many environmental particles simultaneously.

\section{Collapse basis of a flying bullet}

To our knowledge, although there have been many studies about the pointer basis \cite{Kubler1973, Zurek1993a, Zurek1993b, Diosi2000}, none has succeeded in deriving a satisfactory pointer (collapse) basis for a macroscopic flying body in the air. According to everyday observations of objects such as a baseball, the collapse basis of a flying body should have a sharply localized position and a well-defined momentum. Here, we attempt to obtain the desired collapse basis using the entangling-speed-threshold theory. 

We consider the situation illustrated in Fig.~5, where a bullet is flying through the air. The bullet is assumed to be a cubic object with a side length of $2a$ and a mass of $m$. For simplicity, we consider only the one-dimensional center-of-mass degree of freedom $x$. We further assume that the only environmental particles are air molecules. 

\begin{figure}[tb]
        \centering\includegraphics[width=5cm]{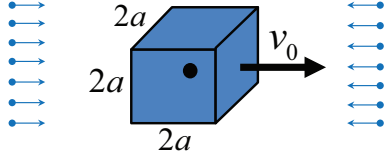}
        \caption{A bullet flying through the air. The bullet is assumed to be a cubic object with a side length of $2a$ and a mass of $m$. We consider only the one-dimensional center-of-mass degree of freedom $x$.}
\end{figure}

The interaction energy function between the bullet and a single molecule is assumed to be a function of the distance between them, as shown in Fig.~6, where ${x_{\cal A}}$ and ${x_{i}}$ are the coordinates of the center of mass of the bullet and the ${i^{{\rm{th}}}}$ molecule, respectively. To quantize the system, we replace ${x_{\cal A}}$ and ${x_{i}}$ with operators ${\hat x_{\cal A}}$ and ${\hat x_{i}}$. Then, the interaction operator ${\hat V_i}$ with the ${i^{{\rm{th}}}}$ molecule should be a function of the ${i^{{\rm{th}}}}$ distance operator ${\hat d_i} \equiv ({\hat x_{\cal A}} - {\hat x_i})$. 

\begin{figure}[tb]
        \centering\includegraphics[width=7cm]{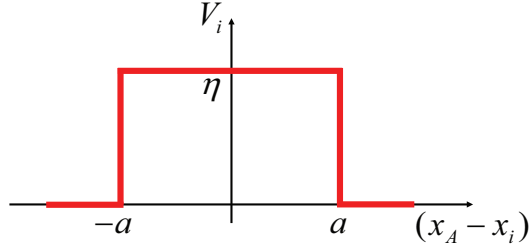}
        \caption{Interaction energy function ${V_i}$ between the bullet and the ${i^{{\rm{th}}}}$ air molecule. ${x_{\cal A}}$ and ${x_i}$ are the coordinates of the center of mass of the bullet and the ${i^{{\rm{th}}}}$ molecule, respectively. A step-function type potential barrier is adopted, with a barrier energy $\eta $.}
\end{figure}

The distance between two objects depends on their relative velocity because of relativistic length contraction. Although we are considering nonrelativistic quantum mechanics and a slowly flying bullet, we will see that relativistic effects should be taken into account when determining the collapse basis because a sufficiently large number of environmental particles causes this otherwise negligible effect to become considerable. If we denote the velocity operator of the bullet by ${\hat v_{\cal A}}$ and that of the ${i^{{\rm{th}}}}$ molecule by ${\hat v_i}$, then the relativistic distance operator $\hat d_i^{({\rm{rel}})}$ of the two objects is given by 
\begin{equation}
\hat d_i^{({\rm{rel}})} = \sqrt {1 - {{{{\left( {{{\hat v}_{\cal A}} - {{\hat v}_i}} \right)}^2}} \over {{c^2}}}} \left( {{{\hat x}_{\cal A}} - {{\hat x}_i}} \right),
\end{equation}
where $c$ is the speed of light. When relativistic effects are considered, the interaction operator ${\hat V_i}$ is a function of $\hat d_i^{({\rm{rel}})}$ instead of ${\hat d_i}$. 

We use the collapse operator $\hat C$ of Eq.~(9) to derive the collapse basis of the flying bullet. Because the interaction Hamiltonian ${\hat H_{{\rm{int}}}}$ is the sum of each ${\hat V_i}$, $\hat C$ is given by
\begin{equation}
\hat C = \langle {{\cal E}_0}|\sum\limits_i {{{\hat V}_i}} \left( {\hat d_i^{({\rm{rel}})}} \right)\left| {{{\cal E}_0}} \right\rangle ,
\end{equation}
where $\left| {{{\cal E}_0}} \right\rangle $ is the typical environmental state that we should infer. We assume that the ${i^{{\rm{th}}}}$ air molecule is in its own state $\left| {e{}_i} \right\rangle $ and that the typical environmental state $\left| {{{\cal E}_0}} \right\rangle $ is a product state as follows: 
\begin{equation}
\left| {{{\cal E}_0}} \right\rangle  = \left| {{e_1}} \right\rangle \left| {{e_2}} \right\rangle  \cdots \left| {{e_N}} \right\rangle .
\end{equation}

The presence of the bullet constrains the possible state of the surrounding air molecules. For example, no air molecules can reside in the region occupied by the bullet. Suppose that the center of mass of the bullet is located at ${x_0}$. Then, it is reasonable to assume that the number density of the air molecules, $n(x)$, is similar to that depicted in Fig.~7. We further assume that the ambient air molecules are in bulk motion at a velocity of ${v_0}$. This bulk motion can arise from two possible causes. The first is wind. The second is the velocity of the bullet itself. Because the air molecules surround the bullet, the ambient molecules must be moving at nearly the same velocity as the bullet. In summary, we specify $\left| {{{\cal E}_0}} \right\rangle $ as a product state such that the air molecules are absent from the region occupied by the bullet and have a velocity of ${v_0}$. 

\begin{figure}[tb]
        \centering\includegraphics[width=7cm]{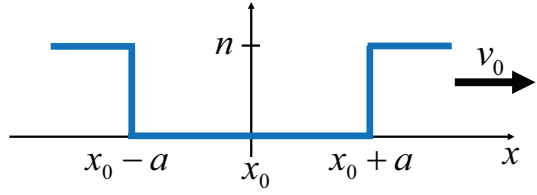}
        \caption{One-dimensional number-density distribution $n(x)$ of the air molecules surrounding the bullet. Air molecules are absent from the region occupied by the bullet, whose center of mass is located at $x_0$. The typical environmental state $\left| {{{\cal E}_0}} \right\rangle $ is defined such that all of the air molecules are in bulk motion at a velocity of $v_0$.}
\end{figure}

Let us sketch our strategy for obtaining an explicit expression for the collapse operator $\hat C$. First, we insert Eq.~(19) into Eq.~(18) to obtain 
\begin{equation}
\hat C = \sum\limits_i {\langle {e_i}|{{\hat V}_i}} \left( {\sqrt {1 - {{{{\left( {{{\hat v}_{\cal A}} - {{\hat v}_i}} \right)}^2}} \over {{c^2}}}} \left( {{{\hat x}_{\cal A}} - {{\hat x}_i}} \right)} \right)\left| {{e_i}} \right\rangle .
\end{equation}
Instead of calculating the ${i^{{\rm{th}}}}$ term individually, we directly consider the result after summing over all individual terms. After the summation, $\hat C$ should be a function of ${\hat x_{\cal A}}$ and ${\hat v_{\cal A}}$. Because we assumed that all air molecules are in bulk motion at velocity $v_0$, if ${\hat v_{\cal A}} = {v_0}$, then there is no relativistic velocity effect. Similarly, if ${\hat x_{\cal A}} = {x_0}$, then no air molecule can have a high interaction energy because of the number-density distribution of Fig.~7. Consequently, $\hat C$ should be equal to zero (at a minimum) when ${\hat v_{\cal A}} = {v_0}$ and ${\hat x_{\cal A}} = {x_0}$. Our strategy is to obtain an approximate formula for $\hat C$ by separately considering these two effects, i.e., the velocity effect and the position effect, near the minimum of $\hat C$. 

Neglecting the relativistic velocity effect for the moment, let us first investigate the summed result of the position term $\left( {{{\hat x}_{\cal A}} - {{\hat x}_i}} \right)$ in Eq.~(20). Figure~8(a) illustrates how this term affects. In the position-basis representation, if ${x_{\cal A}}$ deviates from ${x_0}$, then the air molecules in the shaded region will enter the region of high potential energy $\eta $. The number of air molecules in the shaded region should be $n\left| {{x_{\cal A}} - {x_0}} \right|$. Hence, the summed result of the position term $\left( {{{\hat x}_{\cal A}} - {{\hat x}_i}} \right)$ under the condition ${\hat v_{\cal A}} = {v_0}$ is 
\begin{equation}
{\left. {\hat C} \right|_{{{\hat v}_{\cal A}} = {v_0}}} = \eta n\left| {{{\hat x}_{\cal A}} - {x_0}} \right|.
\end{equation}

\begin{figure}[tb]
        \centering\includegraphics[width=10cm]{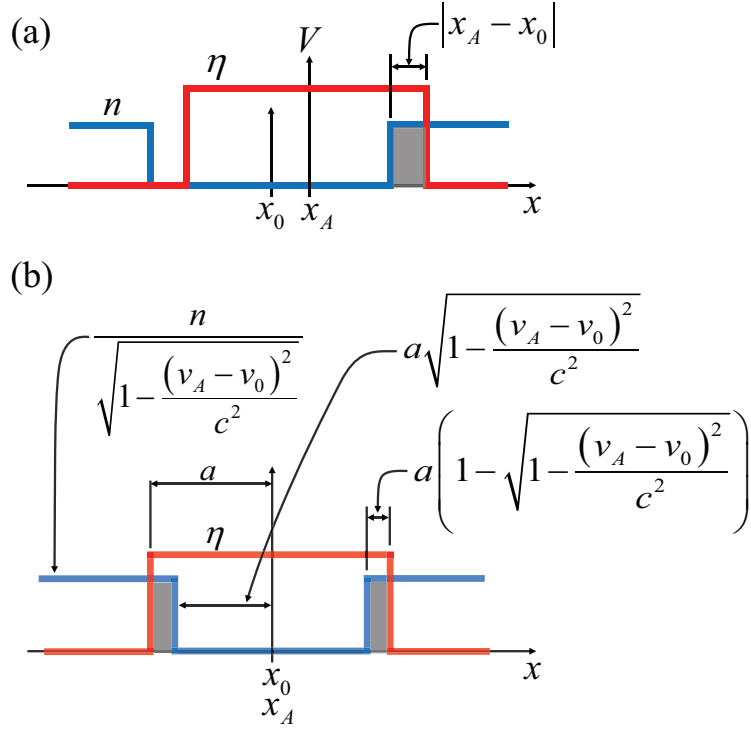}
        \caption{Separation of the position effect and the velocity effect in the collapse operator $\hat C$ of Eq.~(20). (a) Position effect. If the bullet coordinate ${x_{\cal A}}$ were to shift away from ${x_{0}}$, then the air molecules in the shaded region would enter the high-potential-energy region. (b) Velocity effect (${v_{\cal A}}$ and ${v_{0}}$ are the velocities of the bullet and the ambient air, respectively). If the relative velocity $\left( {{v_{\cal A}} - {v_0}} \right)$ is nonzero, then the bullet experiences the air molecules as being closer because of relativistic length contraction. Hence, the air molecules in the shaded region enter the high-potential-energy region. The number density $n$ of the air molecules also increases because of length contraction.}
\end{figure}

Now, we examine the summed result of the velocity term $\left( {{{\hat v}_{\cal A}} - {{\hat v}_i}} \right)$ in Eq.~(20). Even when ${x_{\cal A}} = {x_0}$, some air molecules can enter the high-potential-energy region because of relativistic length contraction, as illustrated in Fig.~8(b). The length of the shaded region is $2a\left( {1 - \sqrt {1 - {{\left( {{v_{\cal A}} - {v_0}} \right)}^2}/{c^2}} } \right)$. We further note that the number density of the air molecules increases because of length contraction such that $n$ is replaced by $n/\sqrt {1 - {{\left( {{v_{\cal A}} - {v_0}} \right)}^2}/{c^2}} $. Hence, the summed result of the term $\left( {{{\hat v}_{\cal A}} - {{\hat v}_i}} \right)$ under the condition ${\hat x_{\cal A}} = {x_0}$ is 
\begin{equation}
{\left. {\hat C} \right|_{{{\hat x}_{\cal A}} = {x_0}}} = 2\eta na{{1 - \sqrt {1 - {{\left( {{{\hat v}_{\cal A}} - {v_0}} \right)}^2}/{c^2}} } \over {\sqrt {1 - {{\left( {{{\hat v}_{\cal A}} - {v_0}} \right)}^2}/{c^2}} }}.
\end{equation}

Near the minimum, $\hat C$ can be expressed as the sum of Eq.~(21) and (22), i.e., 
\begin{equation}
\hat C = \eta n\left| {{{\hat x}_{\cal A}} - {x_0}} \right| + 2\eta na{{1 - \sqrt {1 - {{\left( {{{\hat v}_{\cal A}} - {v_0}} \right)}^2}/{c^2}} } \over {\sqrt {1 - {{\left( {{{\hat v}_{\cal A}} - {v_0}} \right)}^2}/{c^2}} }}.
\end{equation}
After performing a Taylor series expansion of the velocity term up to the first non-zero order, we obtain a simplified formula, 
\begin{equation}
\hat C = {{\eta na} \over {{c^2}}}{\left( {{{\hat v}_{\cal A}} - {v_0}} \right)^2} + \eta n\left| {{{\hat x}_{\cal A}} - {x_0}} \right|.
\end{equation}
Because ${\hat v_{\cal A}} = {\hat p_{\cal A}}/m$ and ${\hat p_{\cal A}} =  - i\hbar (\partial /\partial {x_{\cal A}})$, Eq.~(24) is rewritten in the position-basis representation as follows:
\begin{equation}
\hat C = {{\eta na{\hbar ^2}} \over {{m^2}{c^2}}}\left[ {{{\left( { - i{\partial  \over {\partial x}} - {{{p_0}} \over \hbar }} \right)}^2} + b\left| {x - {x_0}} \right|} \right],
\end{equation}
where
\begin{equation}
b \equiv {{{m^2}{c^2}} \over {a{\hbar ^2}}}
\end{equation}
and ${p_0} = m{v_0}$. Above, we have dropped the suffix ${\cal A}$ for brevity. 

We are now in a position to obtain the collapse basis, i.e., the eigenbasis of $\hat C$. Equation~(25) is the well-known ``vee potential'' Hamiltonian and has analytical eigenfunctions \cite{Blinder2010Http}. We focus only on the state with the lowest-eigenvalue. The lowest eigenfunction is
\begin{equation}
{\psi _0}(x) = {e^{i({p_0}/\hbar )x}}{\rm{Ai}}\left[ {{2^{1/3}}{b^{ - 2/3}}\left( {b\left| {x - {x_0}} \right| - 0.808614 \cdot {b^{2/3}}} \right)} \right],
\end{equation}
where Ai is the Airy function. This Airy function is a localized wave packet at $x_0$ with an average momentum of $p_0$ (i.e., a velocity of $v_0$). Because the size of the wave packet of Eq.~(27) is on the order of ${b^{ - 1/3}}$, the position uncertainty $\Delta x$ is also on the order of ${b^{ - 1/3}}$. If we use the relation $a = (1/2){(\rho /m)^{1/3}}$ (where $\rho $ is the density of the bullet), $\Delta x$ becomes 
\begin{equation}
\Delta x \approx {\left( {{{{\hbar ^2}} \over {2{c^2}{\rho ^{1/3}}}}} \right)^{1/3}} \cdot {m^{ - 5/9}},
\end{equation}
Now, the momentum uncertainty $\Delta p$ can be approximated by the Heisenberg uncertainty relation $\Delta x\Delta p\sim\hbar /2$. Because Eq.~(27) is very similar to a Gaussian wave packet, we can safely use this relation. Instead of $\Delta p$, we prefer to use the more recognizable quantity $\Delta v = \Delta p/m$, which is given by
\begin{equation}
\Delta v \approx {1 \over 2}{\left( {2\hbar {c^2}{\rho ^{1/3}}} \right)^{1/3}} \cdot {m^{ - 4/9}}.
\end{equation}
Hence, both $\Delta x$ and $\Delta v$ decrease as the mass of the bullet increases, indicating that a more massive object has much less uncertainties in both $\Delta x$ and $\Delta v$. This finding is consistent with our intuition from the classical world. Let us estimate specific values of these uncertainties for a steel bullet with a mass of $m = 10{\rm{ g}}$. In that case, $\Delta x = {\rm{1}}.{\rm{9}} \times {\rm{1}}{{\rm{0}}^{ - 28}}~{\rm{ m}}$ and $\Delta v = {\rm{2}}.{\rm{8}} \times {\rm{1}}{{\rm{0}}^{ - 5}}~{\rm{ m/s}}$. Consequently, we have succeeded in showing that the entangling-speed-threshold theory naturally produces a collapse basis for a flying bullet that is consistent with our everyday experience, as desired. 

If we use a more realistic potential function ${V_i}$ in Fig.~5, we arrive at a more plausible collapse basis. For example, if we use a trapezoid-shaped ${V_i}$ in Fig.~5, we obtain a quadratic potential in Eq.~(25) instead of the ``vee potential''. As a result, we obtain the collapse basis in the familiar form of Hermite-Gaussian wave packets instead of the Airy function. 

The method for obtaining the collapse basis of a bullet can also be applied to a spring-mass system (harmonic oscillator) surrounded by air molecules. Because the interaction Hamiltonian with the air molecules is the same for both the bullet and the spring-mass system, the collapse operator $\hat C$ and, consequently, the collapse basis are the same in both cases. The only difference between the flying bullet and the spring-mass system is that the self-Hamiltonians of the two systems are different. This difference gives rise to a difference in the way that the systems unitarily evolve during the time intervals between collapses. Therefore, the dynamics of the two systems appear to be different, although their collapse bases are of the same kind. 

In the entangling-speed-threshold theory, the self-Hamiltonian plays no role in determining the collapse basis; the interaction Hamiltonian and the environmental state are involved, as can be seen in Eq.~(9). In this respect, our theory is different from previous research \cite{Kubler1973, Zurek1993a, Diosi2000}, in which the authors argued that the pointer states of a harmonic oscillator are coherent states that are defined by the self-Hamiltonian.

Notably, the collapse basis of Eq.~(27) does not depend on the potential barrier $\eta $ in Fig.~6 or on the density of the air molecules $n$ in Fig.~7. The uncertainties $\Delta x$ and $\Delta v$ in the collapse states are also independent of $\eta $ and $n$. This independence is because $\eta $ and $n$ became the common multiplication factor in $\hat C$, as can be seen in Eq.~(25). However, this fact does not mean that $\eta $ and $n$ play no role in the collapse process. $\eta $ and $n$ are important to the collapse process for two reasons. First, they determine when collapse occurs. To fulfill the collapse condition of Postulate~1, the entangling speed must reach the threshold, and $\eta $ and $n$ determine how rapidly the entangling speed increases. Second, $\eta $ and $n$ determine how accurately energy is conserved before and after collapse. We argued in Section~VI that energy is well conserved under the condition that the self-Hamiltonian ${\hat H_{\cal A}}$ is negligible compared with the collapse operator $\hat C$. Because ${\hat H_{\cal A}} = (m/2)\hat v_{\cal A}^2$ for the bullet, we can directly compare the coefficients of the $\hat v_{\cal A}^2$ terms in ${\hat H_{\cal A}}$ and in $\hat C$ given in Eq.~(24). $\eta na/{c^2}$ is $3.9 \times {10^4}$ times greater than $m/2$ in a realistic situation characterized by the following values: $\eta  = 1~{\rm{ J}}$, $m = 10~{\rm{ g}}$, $a = 0.54~{\rm{ cm}}$ (calculated from the density of steel),  $n = 3.2 \times {10^{19}}~{\rm{c}}{{\rm{m}}^{ - 1}}$ (inferred from the number density of ideal gases at $0^\circ {\mathrm{C}}$ and 1~atm, and the cross-sectional area of the bullet, $4{a^2}$). Hence, sufficiently large values of $\eta $ and $n$ ensure energy conservation to a high accuracy.

To avoid any misconceptions on the part of the readers, it is worth noting two aspects of our method for determining the collapse basis. First, in Fig.~8, the penetration of air molecules into the bullet is fictitious. It is not an actual penetration but merely a calculation procedure followed by Eq.~(20). Second, our method does not rely on circular logic, even though we assumed the bullet to be initially located at ${x_0}$ with a velocity of ${v_0}$. This assumption would not be necessary if we were given an environmental state $\left| {{{\cal E}_0}} \right\rangle $. In a realistic situation, the most plausible environmental state is one such that air molecules are absent in the region occupied by the bullet and have some velocity due to either wind or the motion of the bullet itself. Thus, what we assumed is the environmental state rather than the state of the bullet. The essential message is that given a plausible environmental state, the collapse basis is derived to have very small uncertainties in both position and momentum.

\section{Experimental verification}

We propose a thought experiment that can verify or falsify the entangling-speed-threshold theory. Although this thought experiment is not feasible at this time, we expect that this proposal can serve as a guide for devising a more realistic scheme in the future. 

We use the spin model considered in Section~IV, with a slight modification. Using the same settings as Section~IV, we modify the Hamiltonian to a simpler one such that 
\begin{equation}
\hat H = {\hat \sigma _{z,{\cal A}}} \otimes g\sum\limits_{k = 1}^N {{{\hat \sigma }_{z,k}}} ,
\end{equation}
where $g$ is the common coupling constant. Suppose that we prepare the initial state as $\left| {\Psi (0)} \right\rangle  = {\left|  +  \right\rangle _{\cal A}} \otimes \left|  +  \right\rangle _{\cal E}^{ \otimes N}$. In this case, there is an analytic solution for an evolved state (using units of $\hbar  = 1$),
\begin{equation}
\left| {\Psi (t)} \right\rangle  = {1 \over {\sqrt 2 }}{\left| 0 \right\rangle _{\cal A}} \otimes \left[ {{1 \over {\sqrt 2 }}\left( {{e^{ - igt}}\left| 0 \right\rangle  + {e^{igt}}\left| 1 \right\rangle } \right)} \right]_{\cal E}^{ \otimes N} + {1 \over {\sqrt 2 }}{\left| 1 \right\rangle _{\cal A}} \otimes \left[ {{1 \over {\sqrt 2 }}\left( {{e^{igt}}\left| 0 \right\rangle  + {e^{ - igt}}\left| 1 \right\rangle } \right)} \right]_{\cal E}^{ \otimes N}.
\end{equation}
As $N$ increases, the entangling speed $\dot \epsilon $ of ${\cal A}$ increases more rapidly and reaches a higher value, as previously observed in Fig.~4 (despite the different Hamiltonians, both systems exhibit the same tendency). Therefore, if we can control the variation of $N$, ${\cal A}$ will begin to satisfy the collapse condition of Postulate~1 at a certain critical number ${N_C}$. 

The question is how to measure whether ${\cal A}$ collapses or not. Our strategy is to wait until the revival time ${t_{{\rm{rev}}}} = 2\pi /g$ at which $\left| {\Psi (t)} \right\rangle $ returns to the initial state. We exploit the fact that the revival time is independent of $N$, whereas the entangling speed reaches a higher value with increasing $N$. At the revival time, ${\cal A}$ should be in the pure state $\left|  +  \right\rangle $. At the time, let us measure ${\cal A}$ in the $\left\{ {\left|  +  \right\rangle ,\left|  -  \right\rangle } \right\}$ basis using a traditional spin-measurement device. If ${\cal A}$ has not collapsed, the result will certainly be $\left|  +  \right\rangle $. However, if ${\cal A}$ has collapsed, the result can occasionally be $\left|  -  \right\rangle $ because a collapse would disrupt the quantum state from its gentle unitary evolution. Hence, as $N$ increases, if we suddenly observe the measurement result $\left|  -  \right\rangle $, we can conclude that ${\cal A}$ collapsed at the corresponding value of $N$. In such an experiment, abrupt observation of a $\left|  -  \right\rangle $ result could be regarded as verification of the entangling-speed-threshold theory. Moreover, the threshold value ${\dot \epsilon_{{\rm{th}}}}$ could be measured in this experiment.

\section{Conclusion}

We proposed the entangling-speed-threshold theory to resolve the quantum measurement problem or, in a more general sense, the problem of the quantum-to-classical transition. The proposed theory is an interaction-induced collapse theory, which is in stark contrast with the spontaneous collapse theory \cite{Ghirardi1986}. The theory consists of two postulates. Postulate~1 states where and when collapse occurs. The question of where collapse occurs is answered by stating that among the many possible bipartite tensor product structures, the single degree of freedom with the maximum entangling speed becomes the epicenter of collapse. The moment when collapse occurs is stated to be the instant when the entangling speed reaches a certain threshold. Postulate~2 states how the collapse basis is determined. The collapse basis is the basis that is most robust against entanglement formation at the instant immediately after collapse. Mathematically, the collapse basis is determined as the eigenbasis of the so-called ``collapse operator''. 

Using the entangling-speed-threshold theory, we can answer many open questions. First, the problem of what the systems are (in other words, what the actual tensor product structure is) is answered by arguing that a system is defined as a degree of freedom that can collapse. Second, the question of how deterministic classical dynamics emerges from indeterministic quantum collapse is explained by nearly continuous collapse causing a certain collapse state to arise with a high Born's probability near 1. Third, observables (or, more generally, measurement operators) are derived by considering how the measured system, the measuring apparatus, and the environment interact. Fourth, we showed that energy is conserved to a high accuracy not only in ensemble level but also in a single collapse event, as long as the environment consists of many degrees of freedom. 

We applied the theory to specific examples to check whether the theory is consistent with known phenomena. Using a spin model, we demonstrated that, as the number of environmental particles increases, the entangling speed reaches a much higher value and energy is better conserved. Next, we considered a bullet flying through the air. We succeeded in deriving the expected collapse basis, which is highly localized in both position and momentum. Finally, we proposed a thought experiment that is capable of verifying the entangling-speed-threshold theory.

One guiding principle that we considered while building the theory was that the collapse condition of Postulate~1 should be Galilei-invariant. Indeed, the entangling speed is Galilei-invariant in every reference frame. To extend the theory to the relativistic regime, the theory should be Lorentz-invariant. To our knowledge, it is not clear whether the entangling speed is Lorentz-invariant. If not, the theory should be modified to be applicable to the relativistic regime.

\begin{acknowledgments}
I appreciate Hee Joon Jeon for helpful discussions. This work was partly supported by the IT R\&D program of MOTIE/KEIT [10043464 (2016)].
\end{acknowledgments}

\bibliography{Bib_Total_Yun}


\end{document}